\documentclass[letterpaper,twocolumn,fleqn]{article}

\usepackage{ist}
\usepackage{cite}
\usepackage{xcolor}

\usepackage{array}
\newcolumntype{P}[1]{>{\centering\arraybackslash}p{#1}}

\usepackage{graphicx,subcaption}
\usepackage{caption}

\usepackage{graphicx,amsmath,amsfonts,amscd,amssymb,bm,url,color,latexsym}
\usepackage{physics}
\allowdisplaybreaks
\usepackage[capitalize,nameinlink]{cleveref}
\usepackage{mathrsfs}

\newtheorem{theorem}{Theorem}

\newtheorem{proposition}[theorem]{Proposition}

\renewcommand{\mathbf}{\boldsymbol}

\usepackage{bm}

\newcommand{\mb}{\mathbf}
\newcommand{\mc}{\mathcal}

\newcommand{\bb}{\mathbb}

\newcommand{\reals}{\bb R}

\newcommand{\R}{\reals}
\newcommand{\Cp}{\bb C}
\newcommand{\Z}{\bb Z}

\newcommand{\paren}{\pqty}

\newcommand{\ol}{\overline}

\usepackage{subcaption}
\usepackage{algorithm}
\usepackage{algpseudocode}
\usepackage{booktabs}
\usepackage{multirow}
\usepackage{dsfont}

\title{What's Wrong with End-to-End Learning for Phase Retrieval?}

\author{Wenjie Zhang$^1$, Yuxiang Wan$^1$, Zhong Zhuang$^2$, Ju Sun$^1$  \\
$^1$Department of Computer Science and Engineering, University of Minnesota, Email: \{zhan7867,wan01530,jusun\}@umn.edu\\
$^2$Department of Physics and Astronomy, University of California, Los Angeles, Email: zhong@physics.ucla.edu
} 
\date{} 

\begin{document} 

\maketitle



\begin{abstract}
For nonlinear inverse problems that are prevalent in imaging science, symmetries in the forward model are common. When data-driven deep learning approaches are used to solve such problems, these intrinsic symmetries can cause substantial learning difficulties. In this paper, we explain how such difficulties arise and, more importantly, how to overcome them by preprocessing the training set before any learning, i.e., symmetry breaking. We take far-field phase retrieval (FFPR), which is central to many areas of scientific imaging, as an example and show that symmetric breaking can substantially improve data-driven learning. We also formulate the mathematical principle of symmetry breaking. 
\end{abstract} 


\section{Introduction}
\begin{figure}[!htbp]
\centering
  \includegraphics[width=1\columnwidth]{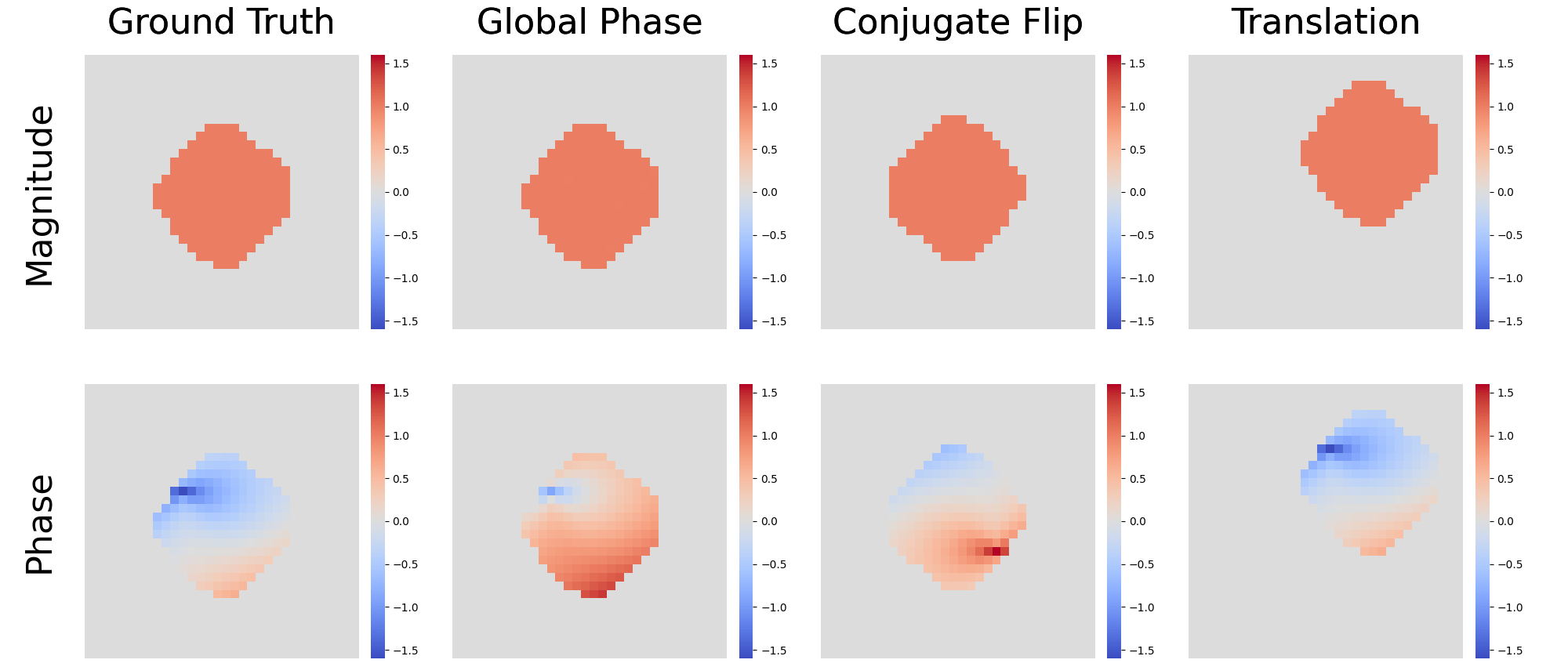}
  \caption{Illustration of the three intrinsic symmetries of FFPR: (i) global phase shift, (ii) 2D conjugate flipping and (iii) translation. Any composition of these three symmetries, when applied to a feasible solution $\mb X$, will lead to the \textbf{same measurement} $\mb Y$ in the Fourier domain. }
\label{fig:symmetries}
\end{figure}

Far-field phase retrieval (FFPR) is a nonlinear inverse problem (IP) that is central to several areas of scientific imaging~\cite{BendoryEtAl2017Fourier, ShechtmanEtAl2015Phase}. Given
$\mb Y = \abs{\mathcal{F}\paren{\mb X}}^2 \in \mathbb{R}_+^{M_1 \times M_2}$ ($\R_+$ means nonnegative reals, $\mathcal{F}$ is the oversampled Fourier operator, and $|\cdot|^2$ takes element-wise squared magnitudes and induces nonlinearity), FFPR is about recovering the complex-valued $\mb X \in \Cp^{N_1 \times N_2}$ from $\mb Y$. Fourier phases are lost in the measurement process because detectors in practical imaging systems cannot record complex phases. To ensure recoverability, $M_1 \ge 2N_1 -1$ and $M_2 \ge 2N_2 -1$ are necessary~\cite{Hayes1982reconstruction, jaganathan2015phase}. Under such recoverability condition, FFPR is almost everywhere injective, up to three intrinsic symmetries---operations on $\mb X$ that leave $\mb Y$ unchanged: translation, conjugate flipping, and global phase; see \cref{fig:symmetries}. 

In standard practice, FFPR is solved by meticulously designed iterative numerical methods, e.g., the famous hybrid input-output (HIO) method~\cite{Fienup1982Phase}, in combination with the shrinkwrap heuristic~\cite{marchesini2003x}. The advent of data-driven deep learning (DL) has created numerous opportunities for novel methods for FFPR. For example, in the end-to-end DL approach, training sets of the form $\{(\mb Y_i, \mb X_i)\}$ are used to train deep neural networks (DNNs) that predict $\mb X$ given $\mb Y$~\cite{GoyEtAl2018Low, CherukaraEtAl2018Real, UelwerEtAl2019Phase, MetzlerEtAl2020Deep,Kang:20, Ye_2022}. Recent work~\cite{Zhang:21,ma2022admm} has used pre-trained deep-generative models as plug-in priors to improve recovery performance. 

In this paper, we highlight an overlooked learning difficulty associated with the end-to-end approach to FFPR. The difficulty is caused by the three intrinsic forward symmetries (\cref{fig:symmetries}). As we illustrate later, these symmetries cause the inverse function that DNNs try to approximate in the end-to-end approach to be highly oscillatory. Such irregularities cannot be effectively learned by existing DNNs learning techniques. To temper the difficulty, we propose a novel technique to preprocess the training set before training, which we call \emph{symmetry breaking}. We show that the proposed technique can substantially improve end-to-end learning performance, regardless of the DNN models used. Preliminary versions of the work have appeared in the workshop paper~\cite{tayal2020unlocking} and a preprint~\cite{tayal2020inverse}.

\begin{figure*}[!htbp]
\centering
\resizebox{1\linewidth}{!}{%
\begin{tabular}{c c c}
\centering
\includegraphics[width=0.32\textwidth]{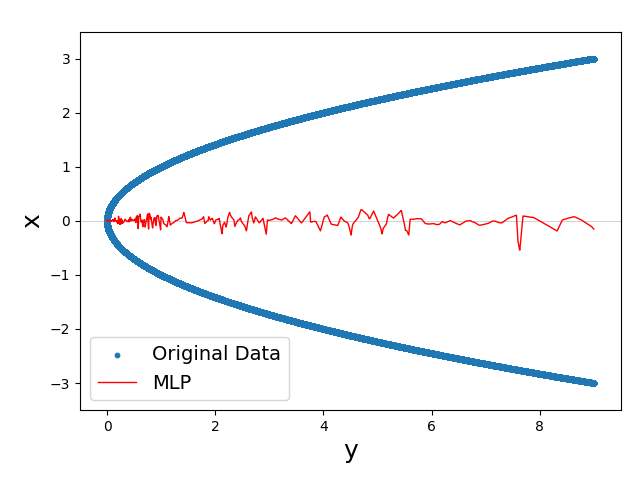}
&\includegraphics[width=0.32\textwidth]{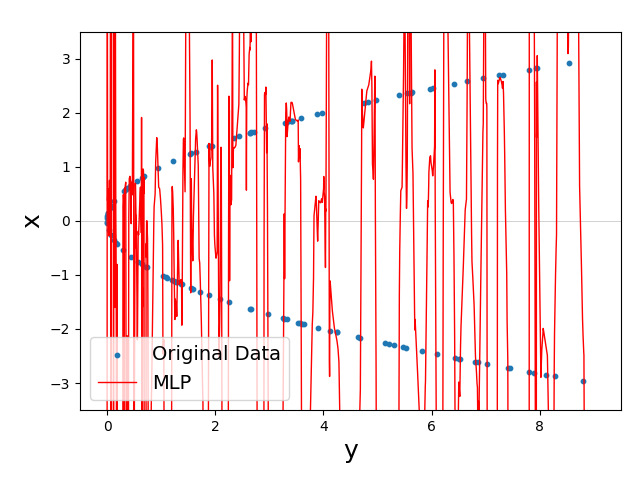}
&\includegraphics[width=0.32\textwidth]{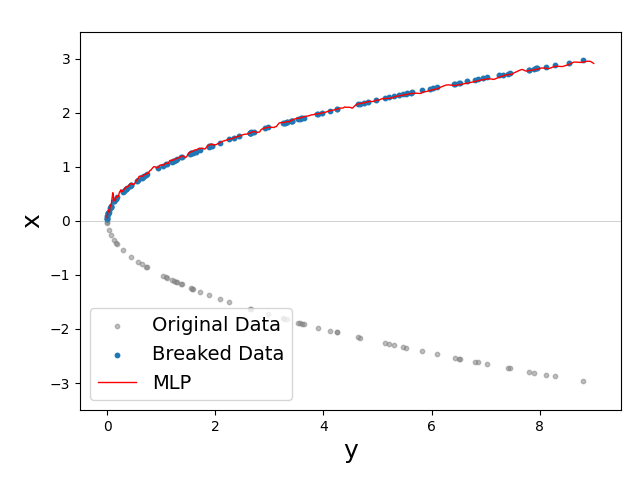}
\\
{\textbf{(i)}}
&{\textbf{(ii)}}
&{\textbf{(iii)}}
\\
\includegraphics[width=0.32\textwidth]{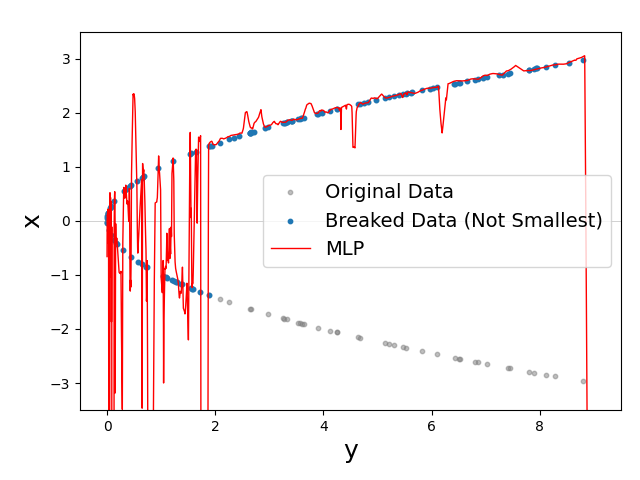}
&\includegraphics[width=0.32\textwidth]{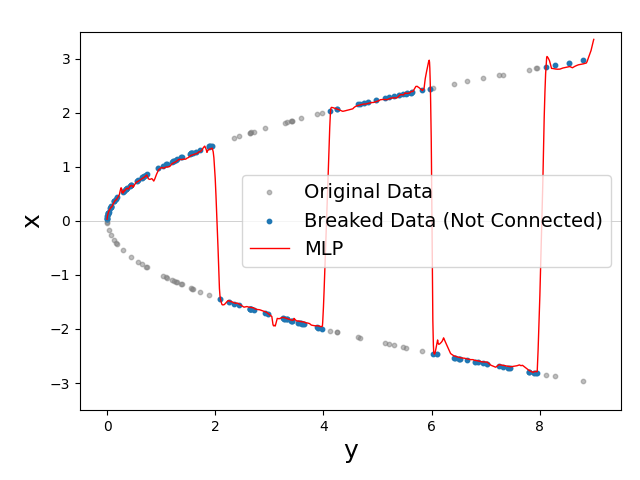}
&\includegraphics[width=0.32\textwidth]{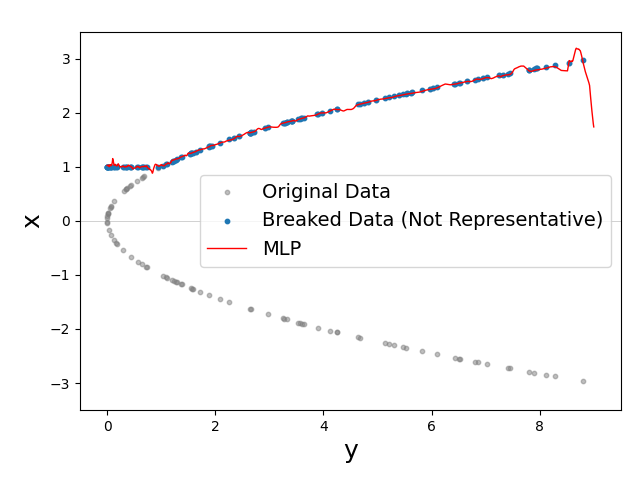}
\\
{\textbf{(iv)}}
&{\textbf{(v)}}
&{\textbf{(vi)}}
\\
\end{tabular}%
}
\caption{Our proposed \emph{symmetry breaking} process substantially improves end-to-end learning on the square root problem. \textbf{Top row}: learning results on the raw training set with ultra-dense sampling (i) and dense sampling (ii), vs. on the training set after symmetry breaking; \textbf{Bottom row}: illustration of the three desired properties after symmetry breaking, without which the end-to-end learning can still suffer: (iv) non-smallest, (v) non- connected, and (vi) non-representative.} 
\label{fig:square_root}
\vskip -0.1in
\end{figure*}

\section{Methods}
\label{sec:methods}
\textbf{Symmetry breaking for learning square root} \quad 
To see why symmetries can cause learning difficulties, consider a simple IP: given $y = x^2$, recover $x$. In other words, we try to learn the square-root function, allowing both positive and negative outputs. To implement the end-to-end approach, we draw random numbers $x_i$'s uniformly from $[-3, 3]$, and construct the training set $\{x_i^2, x_i\}$. We then train a $6$-layer multilayer perceptron (MLP), with $100$ hidden nodes in each layer to ensure sufficient capacity, and with layer-wise batch normalization to ensure training stability. The functions learned by the MLP are shown in \cref{fig:square_root}(i) and \cref{fig:square_root}(ii), on ultra-dense and dense training sets, respectively. In neither case can we learn a function with reasonable generalization. In particular, for the former case, we learn an almost trivial function. Why more is less here? 

The culprit is the intrinsic sign symmetry $\{+x, -x\} \mapsto x^2$. Imagine that if we had sampled every point in $[-3, 3]$ when constructing the training set, the ``function'' induced by the training set would be $x^2 \mapsto \{+x, -x\}$, where the output is an equivalence class induced by the sign symmetry. Practically, we only have a finite training set, with random sampling we are unlikely to see both $(x^2, +x)$ and $(x^2, -x)$ in the training set for any $x \in [-3, 3]$. But we can see $(x_1^2, x_1)$ and $(x_2^2, -x_2)$ in the training set where $x_1, x_2 \ge 0$, and $x_1, x_2$ are close. This implies that due to intrinsic sign symmetry, we can see, perhaps in many cases of the training set, that $y_1$ and $y_2$ are close, but their desired outputs differ in sign and hence are far away. Such a close-inputs-distant-outputs property implies that the target function determined by the training set is oscillatory, at least locally. In particular, the larger the training set, the more oscillatory the function can be. In the extreme case, any given DNN gives up on fitting the data due to the high oscillations, i.e., what we have seen in \cref{fig:square_root}(i). 

Taming the difficulty is easy: we can process the training set into $\{x_i^2, \abs{x_i}\}$ (or $\{x_i^2, -\abs{x_i}\}$), i.e., fixing the sign of the outputs. We call this a \emph{symmetry breaking} process. Symmetry breaking leads to a much smoother target function, and hence much easier for the DNN to learn (see \cref{fig:square_root}(iii)). 

Although the symmetry-breaking process seems natural, there is a principle behind: imagine the continuous case again where we sample every point in $[-3, 3]$. The training set after the symmetry breaking is \emph{representative}---every point in the output space is represented up to the sign symmetry, \emph{smallest}---we do not have more than necessary data points in the processed training set, and geometrically \emph{connected}---to minimize potential oscillation in the target function. The three principles are illustrated in \cref{fig:square_root}(iv)-(vi).

\vspace{1em}
\textbf{Symmetry breaking for phase retrieval} \quad 
A crucial question now is how to derive an effective symmetry-breaking algorithm for FFPR, which has three symmetries. First of all, just to be sure, these symmetries can cause similar oscillation issues like our square root example: consider \cref{fig:symmetries}(iii), i.e., the translated copy. The distance between its 2D conjugate flip version and itself is large, although they have exactly the same set of Fourier magnitudes. A perturbation argument like we did in the square-root example implies that we can easily face a similar close-inputs-distant-outputs issue. 

Breaking the symmetries in the object domain seems hopeless: conjugate 2D flipping and non-zero content translation induce irregular equivalent sets in the object space that are hard to represent. Fortunately, the three symmetries can be equivalently represented in terms of the complex phase $e^{i\mb \theta}$ in the Fourier domain. Let $\mc X$ denote the oversampled Fourier transform of $\mb X$.  Now (1) for 2D translation, any allowable\footnote{The nonzero content cannot translate outside the boundaries.} 2D translation $t_1, t_2\in \Z$ induces the change $\mc X\paren{k_1, k_2} \mapsto \exp[i2\pi ({k_1t_1}/{M_1} + {k_2t_2}/{M_2})] \mc X\paren{k_1, k_2}$; (2) conjugate 2D flipping induces the change $\mc X \mapsto \ol{\mc X}$, i.e., change to the complex phase $e^{i \mb \theta} \mapsto e^{-i \mb \theta}$; and (3) global phase transfer induces the change $\mc X \mapsto e^{i \theta} \mc X$. 
The change due to (2) is a global sign flipping in the \emph{angle} space of $\Theta$, and due to (3) is a line shift in the \emph{angle} space of $\Theta$. The equivalent sets are easy to represent in the angle space, but we take an equivalent representation in the complex phase space to avoid the tricky issue of dealing with the $2\pi$ periodicity in the angle space. However, finding a way to represent the equivalent sets of (1) is still tricky, whether considering it from the angle or from the phase space. 

So our overall strategy is a hybrid of \emph{rigorous} symmetry breaking for global phase shift and 2D conjugate flip in the complex phase space, and \emph{heuristic} ``symmetry breaking'' for translation in the original space--our later experiments show that this combination is effective. To break the 2D translation, we propose simply centering the non-zero content as a heuristic. To break global phase shift and 2D conjugate flip, we perform the geometric construction in the angle space and then translate it back to the phase-space representation. To save space, we omit the intuition behind the construction and present the results directly as follows.

Consider the following set in the phase domain $\bb S^{M_1 \times M_2}$ (the space of $e^{i\mb \theta}$), where $\bb S$ denotes the 1D complex circle: 
\begin{equation}
\centering
\label{eq:rep_fpr}
\begin{split}
    \mathscr{H} \doteq \{ &\mb \Omega \in \Cp^{M_1 \times M_2} ~ : ~\mb \Omega(1,1) = 1,\\
    & \mb \Omega(1,2) \in \bb S_{+}, \mb \Omega(i, j) \in \bb S \; \forall\; \text{other index}\;  \paren{i, j}\},
\end{split}
\end{equation}
Here, $\bb S_+$ the upper half circle.  Formally, $\mathscr{H}$ can be understood as a set
\begin{equation}
\begin{bmatrix}
\{1\} & \bb S_{+} & \bb S & \cdots & \bb S \\
\bb S & \bb S & \bb S & \cdots & \bb S \\
\vdots &  \vdots & \vdots & \ddots & \vdots \\
\bb S &  \bb S& \bb S & \cdots  & \bb S
\end{bmatrix}_{M_1 \times M_2}.
\end{equation}
We can prove the following: 
\begin{proposition}\label{thm:2D}
Consider the conjugate flipping and global phase transfer symmetries only.  The set $\mc H$, except for its negligible subset ${\cal{N}} = \{1 \} \times \{\omega \in \bb S: \textup{Im}\paren{\omega}=0\}^{M_1 \times M_2-1}$, is a connected, smallest representative in the phase domain $\bb S$. 
\end{proposition}
We work with end-to-end DNNs that directly predict the $N_1 \times N_2$ target in the object domain. To perform symmetry breaking, we first center the nonzero content inside $\mb X_j$'s in the training set and then take the oversampled Fourier transform and perform the \textit{symmetry breaking} as implied by Proposition~\ref{thm:2D} in the complex phase space. For any phase matrix $\mb \Omega$, the \textit{symmetry breaking} goes naturally as follows: first a global phase transfer is performed to make $\mb \Omega\paren{1, 1} = 1$, and then a global angle negation is performed, i.e., $\theta \mapsto -\theta$ if the second angle is negative. Here, we assume the angle has been transferred to the range of $(-\pi, \pi]$. We summarize the whole pipeline in \cref{alg:fpr}.
\begin{algorithm}
\caption{Symmetry breaking for FFPR on the training set}\label{alg:fpr}
 \begin{algorithmic}[1]
 \renewcommand{\algorithmicrequire}{\textbf{Input:}}
 \renewcommand{\algorithmicensure}{\textbf{Output:}}
\Require Original training set $\{\mb X_j\}_{j \in \mc J} \subset  \Cp^{N_1 \times N_2}$
\Ensure Training set after symmetry breaking  
\For{$j \in \mc J$}
\State center the nonzero content inside $\mb X_j$ 
\State obtain the oversampled Fourier transform of $\mb X_j$: $\mc X_j = \mc F(\mb X_j)$
\State write $\mc X_j$ in elementwise polar form: $\mc X_j = \abs{\mc X_j}e^{i \mb \Theta}$ and denote $\mb \Omega \doteq e^{i \mb \Theta}$
\State perform global phase transfer to make $\mb \Omega\paren{1, 1} = 1$: $\mb \Omega\paren{k_1, k_2} \gets \overline{\mb \Omega\paren{1, 1}} \mb \Omega\paren{k_1, k_2}$
\If{$\mb \Omega(1,2) \in \bb S_{+}$}
    \State $\mb \Omega\paren{k_1, k_2} \gets \mb \Omega\paren{k_1, k_2}$
\ElsIf{$\mb \Omega(1,2) \notin \bb S_{+}$}
    \State $\mb \Omega\paren{k_1, k_2} \gets \overline{\mb \Omega\paren{k_1, k_2}}$
\EndIf
\State form the new $\mb X_j$: $\mb X_j = \mc F^{-1}(\mc X_j)$
\EndFor
\State form and return the new training set $\{(|\mc F (\mb X_j)|^2, \mb X_j) \}$
\end{algorithmic}
\end{algorithm}
\begin{figure*}[!htbp]
\centering
  \includegraphics[width=\textwidth]{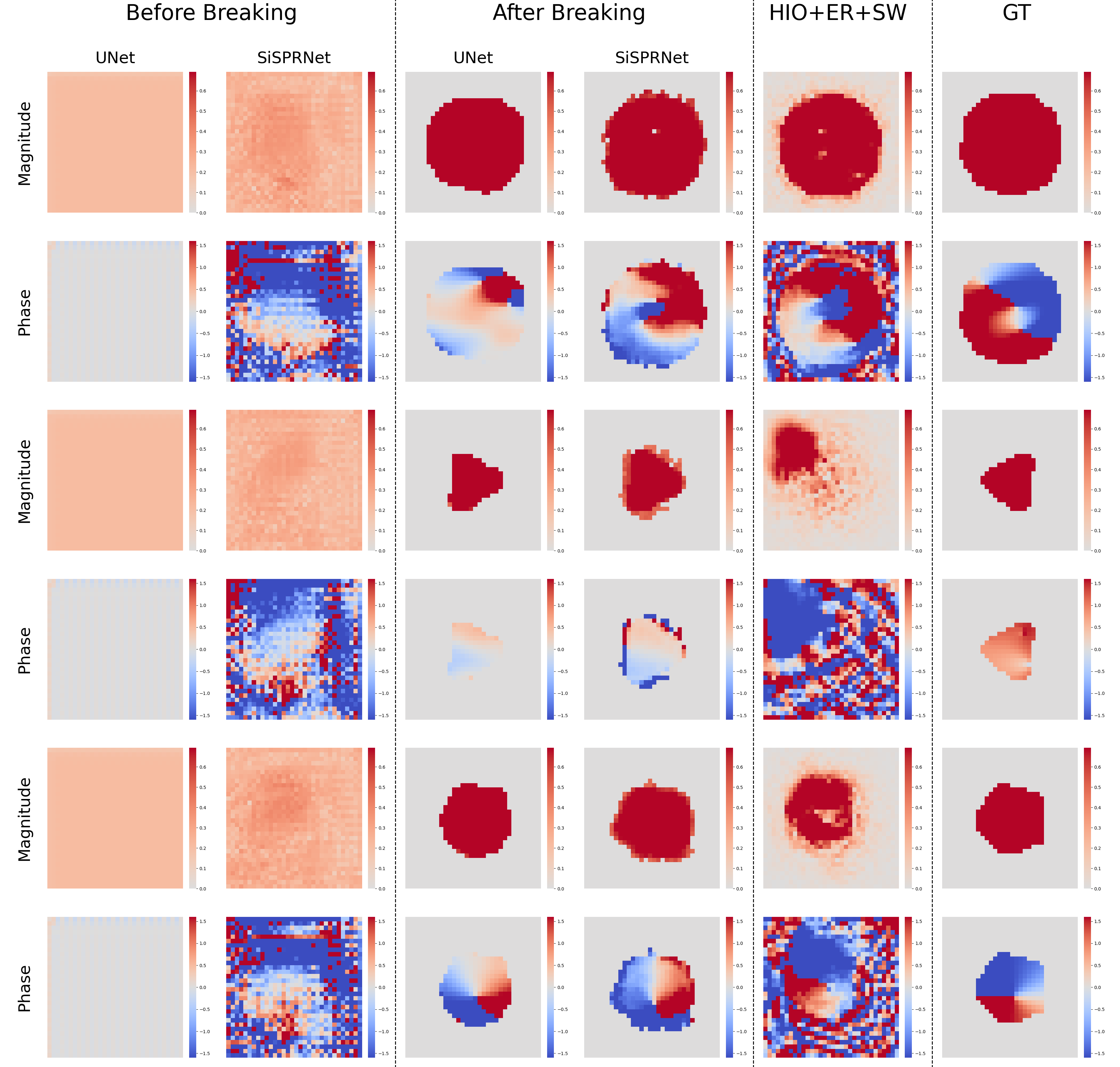}
  \caption{Visual comparison of reconstruction results by different methods. Columns 1 and 2 are the results for UNet and SiSPRNet trained on $5000$ data points before symmetry breaking, columns 3 and 4 are the corresponding results after symmetry breaking. For better visualization of the magnitudes before symmetry breaking, we transform the numbers by the $(\cdot)^{1/4}$ function.}
  \label{Figure:results}
\end{figure*}
\vspace{1em}
\begin{table}[!htbp]
\caption{Reconstruction quality (measured by the optimization/MSE loss) on the \textbf{training set} using the \textbf{model obtained at the final training epoch}. The models are trained on $500$, $1000$, $1500$,  and $5000$ data points, respectively, both before and after \textit{symmetry breaking}. }
\label{tab:NN_far_train_mse_final}
\setlength{\tabcolsep}{1pt}
\renewcommand{\arraystretch}{1}
\centering 
{\small
\begin{tabular}{c|P{1.7cm}P{1.7cm}|P{1.7cm}P{1.7cm}}
\hline
& \multicolumn{2}{c|}{\textbf{UNet}} & 
\multicolumn{2}{c}{\textbf{SiSPRNet}} \\ \hline
\textbf{Sample}  & \textbf{Before}& \textbf{After}& \textbf{Before}& \textbf{After}\\ \specialrule{1pt}{0pt}{0pt}
500 & 0.0396  & 0.0081  & 0.1364 & 0.0106  \\ \hline
1000  & 0.1323  & 0.0075 & 0.1369 & 0.0099  \\ \hline
1500 & 0.1387 & 0.0045 & 0.1387 &  0.0091 \\ \hline
5000  &  0.1390 & 0.0015 & 0.1389 &  0.0071\\ \hline
\end{tabular}
}
\end{table}
\begin{table}[!htbp]
\caption{Reconstruction quality (measured by the SA-MSE loss) on the \textbf{training set} using the \textbf{model obtained at the final training epoch}. The models are trained on $500$, $1000$, $1500$, and $5000$ data points, respectively, both before and after \textit{symmetry breaking}. The SA-MSE loss for HIO+ER+Shrinkwrap is \textbf{0.0449}.
}
\label{tab:NN_far_train_final}
\setlength{\tabcolsep}{1pt}
\renewcommand{\arraystretch}{1}
\centering 
{\small
\begin{tabular}{c|P{1.7cm}P{1.7cm}|P{1.7cm}P{1.7cm}}
\hline
& \multicolumn{2}{c|}{\textbf{UNet}} & 
\multicolumn{2}{c}{\textbf{SiSPRNet}} \\ \hline
\textbf{Sample}  & \textbf{Before}& \textbf{After}& \textbf{Before}& \textbf{After}\\ \specialrule{1pt}{0pt}{0pt}
500 & 0.0707 & 0.0208 & 0.3287  & 0.02734 \\ \hline
1000 & 0.2059 & 0.0186 & 0.3328 & 0.0255  \\ \hline
1500 & 0.2509 & 0.0113 & 0.3310 & 0.0235 \\ \hline
5000 & 0.2520 & 0.0038 & 0.2485 & 0.0183\\ \hline
\end{tabular}
}
\end{table}
\begin{table}[!htbp]
\caption{Reconstruction quality (measured by the optimization/MSE loss) on the \textbf{training set} using the \textbf{model with the smallest validation loss}. The models are trained on $500$, $1000$, $1500$,  and $5000$ data points, respectively, both before and after \textit{symmetry breaking}. }
\label{tab:NN_far_train_mse_best}
\setlength{\tabcolsep}{1pt}
\renewcommand{\arraystretch}{1}
\centering 
{\small
\begin{tabular}{c|P{1.7cm}P{1.7cm}|P{1.7cm}P{1.7cm}}
\hline
& \multicolumn{2}{c|}{\textbf{UNet}} & 
\multicolumn{2}{c}{\textbf{SiSPRNet}} \\ \hline
\textbf{Sample}  & \textbf{Before}& \textbf{After}& \textbf{Before}& \textbf{After}\\ \specialrule{1pt}{0pt}{0pt}
500 & 0.0537 & 0.0119 & 0.2511 & 0.0108  \\ \hline
1000  & 0.1368  & 0.0101 & 0.3064 & 0.0099  \\ \hline
1500 &  0.1387 & 0.0099 & 0.1387 & 0.0095 \\ \hline
5000  & 0.1390 & 0.0089 & 0.1390 &  0.0071\\ \hline
\end{tabular}
}
\end{table}
\begin{table}[!htbp]
\caption{Reconstruction quality (measured by the SA-MSE loss) on the \textbf{training set} using the \textbf{model with the smallest validation loss}. The models are trained on $500$, $1000$, $1500$,  and $5000$ data points, respectively, both before and after \textit{symmetry breaking}. The SA-MSE loss for HIO+ER+Shrinkwrap is \textbf{0.0449}}
\label{tab:NN_far_train_best}
\setlength{\tabcolsep}{1pt}
\renewcommand{\arraystretch}{1}
\centering 
{\small
\begin{tabular}{c|P{1.7cm}P{1.7cm}|P{1.7cm}P{1.7cm}}
\hline
& \multicolumn{2}{c|}{\textbf{UNet}} & 
\multicolumn{2}{c}{\textbf{SiSPRNet}} \\ \hline
\textbf{Sample}  & \textbf{Before}& \textbf{After}& \textbf{Before}& \textbf{After}\\ \specialrule{1pt}{0pt}{0pt}
500 & 0.0857 & 0.0294 & 0.2978 & 0.0278 \\ \hline
1000 & 0.2078 & 0.0243 & 0.3231 & 0.0255 \\ \hline
1500 & 0.2503  & 0.0238 & 0.3324 & 0.0245\\ \hline
5000 & 0.2480 & 0.0208 & 0.2166 & 0.0183\\ \hline
\end{tabular}
}
\end{table}
\begin{table}[!htbp]
\caption{Reconstruction quality (measured by the SA-MSE loss) on the \textbf{test set} using the \textbf{model with the smallest validation loss}. The models are trained on $500$, $1000$, $1500$, and $5000$ data points, respectively, both before and after \textit{symmetry breaking}. The SA-MSE loss for HIO+ER+Shrinkwrap is $0.0441$.}
\label{tab:NN_far_test}
\setlength{\tabcolsep}{1pt}
\renewcommand{\arraystretch}{1}
\centering 
{\small
\begin{tabular}{c|P{1.7cm}P{1.7cm}|P{1.7cm}P{1.7cm}}
\hline
& \multicolumn{2}{c|}{\textbf{UNet}} & 
\multicolumn{2}{c}{\textbf{SiSPRNet}} \\ \hline
\textbf{Sample}  & \textbf{Before}& \textbf{After}& \textbf{Before}& \textbf{After}\\ \specialrule{1pt}{0pt}{0pt}
500 & 0.1393 & 0.0513 & 0.3206 & 0.0458\\ \hline
1000 & 0.2103 & 0.0478 & 0.3246 & 0.0405\\ \hline
1500 & 0.2483 & 0.0465 & 0.3244 & 0.0355\\ \hline
5000 & 0.2496 & 0.0407 & 0.2511 & 0.0289\\ \hline
\end{tabular}
}
\end{table}
\section{Experiments}\label{sc:experiments}

\textbf{Evaluation dataset} \quad 
We use a simulated Bragg CDI (BCDI) crystal dataset as described in~\cite{zhuang2022practical} to construct our training, validation, and test sets. The simulation steps are: (1) Random convex and nonconvex rounded polygons are generated; (2) Magnitudes inside the polygons are set to $1$, and those outside set to $0$, to simulate the uniform magnitudes in practical crystal samples; (3) Complex phases inside the polygons are derived by first placing randomly distributed crystal defects and then projecting the resulting two-dimensional displacement fields onto the corresponding momentum transfer vectors. A variety of polygon shapes and defect densities are included in the dataset to ensure its diversity. The complex-valued images are placed in $128 \times 128$ black backgrounds; see the last column of \cref{Figure:results} for samples. 

In our experiment, we resize the complex-valued images into $32\times 32$ to save computation, and then perform 2$\times$ oversampled Fourier transforms to generate the phaseless measurements. We generate $4$ training sets of varying sizes: $500$, $1000$, $1500$ and $5000$. Moreover, we independently generate a fixed $500$-sample validation set and a $500$-sample test set. The original training sets are designated as ``before breaking'', and those after the symmetry-breaking preprocessing following \cref{alg:fpr} are designated as ``after breaking''.  

\vspace{1em}
\textbf{Experiment setup} \quad 
We choose two DNN backbones, UNet~\cite{CherukaraEtAl2018Real} and SiSPRNet~\cite{Ye_2022} that have been used in end-to-end methods for FFPR. Since our argument about learning difficulty is about the training set and also our symmetry breaking is only performed on the training set, we expect uniform improvement in performance due to the symmetry breaking, regardless of the DNN backbones used. To make these models compatible with our image resolution, we adjust the number of encoder- and decoder-layers in both models. We train both models using the standard MSE (i.e., mean squared error) loss. To evaluate recovery quality, we use the symmetry-adjusted MSE (SA-MSE) as defined in~\cite{tayal2020unlocking}, a modified MSE loss that is invariant to the three intrinsic symmetries. During the training process, we use the Adam optimizer with an initial learning rate of $0.001$, and apply \texttt{ReduceLROnPlateau} which adaptively decreases the learning rate by monitoring the SA-MSE loss on the validation set. We terminate the training process when the learning rate reaches $\le 10^{-5}$. For a baseline method, we pick HIO+ER+Shrinkwrap~\cite{marchesini2003x}, a gold-standard method for FFPR.  

\vspace{1em}
\textbf{Results} \quad 
We present the results from three settings: (1) MSE (\cref{tab:NN_far_train_mse_final}) and SA-MSE  (\cref{tab:NN_far_train_final}) evaluated on the training set using the final models, (2) MSE (\cref{tab:NN_far_train_mse_best}) and SA-MSE (\cref{tab:NN_far_train_best}) evaluated on the training set with the best validated model, and (3) SA-MSE evaluated on the test set with the best validated model (\cref{tab:NN_far_test}).






 
First, we examine how symmetries affect training performance. From \cref{tab:NN_far_train_mse_final,tab:NN_far_train_final}, we observe that symmetry breaking uniformly improves the MSE and SA-MSE loss across the two backbone models, often by an order of magnitude. Moreover, measured by the optimization/MSE loss, it is evident from \cref{tab:NN_far_train_mse_final} that before symmetry breaking the reconstruction quality worsens as the training set grows, while after symmetry breaking the trend is the opposite. All of these observations confirm the learning difficulty and trend that we predict from the square-root example; see \cref{fig:square_root}. Note that the trend measured by SA-MSE is mostly consistent, but not entirely. Presumably, this is due to the potential different behaviors between MSE and SA-MSE. Moreover, in terms of SA-MSE loss, training on the unprocessed training set often performs even worse than the non-data-driven baseline HIO+ER+Shrinkwrap, despite the large capacity of the backbone models we choose. This suggests that symmetry breaking is a crucial step in maximizing learning efficiency for data-driven FFPR.  

From \cref{tab:NN_far_train_mse_best,tab:NN_far_train_best} we again observe the uniform improvement of reconstruction quality due to symmetry breaking on both backbone models. The trend with respect to the size of the training set becomes less clear now, as our selection of the best model with respect to the validation set starts to involve generalization performance---this cannot be fully predicted by our argument. 

\cref{tab:NN_far_test} suggests that the performance boost due to symmetry breaking on the training sets carries on to their respective test sets, i.e., test performance is uniformly improved after symmetry breaking. Note that without symmetry breaking, the test performance is always worse than that of non-data-driven HIO+ER+Shrinkwrap baseline, whereas after symmetry breaking, we can start to see the advantage of data-driven methods: they outperform the baseline method.   

\cref{Figure:results} visually compares the recovery results before and after symmetry breaking on several randomly chosen test samples. Before symmetry breaking, both backbone models typically fail to recover the objects and only find trivial solutions---the reconstructed objects have negligibly small magnitudes. After symmetry breaking, for all of the visualized samples, we observe reasonable magnitude recovery, especially of the boundary details. The estimated phases also reveal relatively sharp details, again confirming the expected performance benefits of symmetry breaking. 
\section{Related work}
\label{sc:related_works_ei}
Recently, there have been intensive research efforts on solving IPs using DL~\cite{McCannEtAl2017Convolutional, LucasEtAl2018Using, ArridgeEtAl2019Solving, Zhang2023kbnet}. The end-to-end approach is attractive not only because of its simplicity but also because (i) we do not even need to know the forward models, so long as we can gather sufficiently many data samples and weak system properties such as symmetries---e.g., this is handy for complex imaging systems~\cite{HorisakiEtAl2016Learning, LiEtAl2018Deep};
(ii) alternatives have rarely worked, and a good example is FFPR that we study in this work~\cite{Fienup1982Phase, SinhaEtAl2017Lensless}.

Besides these, the end-to-end DL approach has been empirically applied to a number of problems with symmetries, e.g., blind image deblurring (i.e., blind deconvolution)~\cite{TaoEtAl2018Scale, Fang_2023_CVPR}, real-valued Fourier phase retrieval~\cite{SinhaEtAl2017Lensless}, 3D surface tangents and normal prediction~\cite{huang2019framenet, Do2022EgocentricSU}, nonrigid structure-from-motion~\cite{kong2019deep, WangEtAl2020Deep}, IPs involving partial differential equations ~\cite{RAISSI2019686,lu2021physicsinformed,KHARAZMI2021113547}, blind source separation ~\cite{electronics12183954}, single image depth estimation. The difficulty of learning highly oscillatory functions by DNNs is also found in DL for solving partial differential equations~\cite{oommen2022learning,adcock2021gap}. 

For FFPR, recent work has also tried to integrate DL modules with the traditional regularized data-fitting framework~\cite{metzler2018prdeep, IsilEtAl2019Deep}. However, HIO is still needed to produce good initialization, and their methods mostly perform only local refinement.  






\section{Acknowledgments} 
W. Zhang and Y. Wan are partially supported by a UMN CSE InterS\&Ections Seed Grant. Dr. Kshitij Tayal, Dr. Jesse Lai and Raunak Manekar were involved in the early development of the ideas included here, as reported in the preliminary work~\cite{tayal2020inverse,tayal2020unlocking}. 



\small
\bibliographystyle{plain}
\bibliography{ref}

\begin{biography}

Wenjie Zhang is a first-year Ph.D. student in Computer Science, University of Minnesota. His research focuses mainly on computer vision and AI for science.

Yuxiang Wan is a research associate at the University of Minnesota. His research focuses mainly on scientific machine learning. 

Zhong Zhuang is an AI for Science Postdoctoral Researcher at UCLA. His research interest is in developing foundational AI tools to advance scientific discovery. Before, he received his PhD degree from Electrical and Computer Engineering of University of Minnesota in 2023. 

Ju Sun is an assistant professor in Computer Science \& Engineering, University of Minnesota. His research interest lies at the theoretical and computational foundations of AI, and AI application in science, engineering, and medicine. Before, he received his PhD degree in electrical engineering from Columbia University (2011--2016) and worked as a postdoctoral scholar at Stanford University (2016--2019). 
\end{biography}
\end{document}